\documentclass[prl,twocolumn,showpacs,hyperref,balancelastpage]{revtex4}%
\usepackage{epsfig}
\usepackage{color}
\usepackage{amsmath}
\usepackage{amsfonts}
\usepackage{amssymb}
\usepackage{graphicx}%
\newcommand{\be}{\begin{equation}}
\newcommand{\ee}{\end{equation}}
\newcommand{\bea}{\begin{eqnarray}}
\newcommand{\eea}{\end{eqnarray}}

\begin{document}
\title{Le Chatelier principle for out of equilibrium and boundary driven systems : application to   dynamical phase transitions. }
\author{O. Shpielberg $^{1}$ and E. Akkermans$^1$}
\affiliation{$^1$Department of Physics, Technion Israel Institute of Technology, Haifa 32000, Israel}

\begin{abstract}

A stability analysis of out of equilibrium and boundary driven systems is presented. It is performed in the framework of the hydrodynamic macroscopic fluctuation theory and assuming the additivity principle whose interpretation is discussed with the help of a Hamiltonian description. An extension of Le Chatelier principle for out of equilibrium situations is presented which allows to formulate the conditions of validity of the additivity principle. Examples of application of these results in the realm of classical and quantum systems are provided. 
\end{abstract}


\date{\today}
\maketitle

Understanding the behaviour of out of equilibrium systems is an essential problem in physics \citep{Redner} but surprisingly enough and except for few exact solutions, it still lacks both a macroscopic approach comparable to thermodynamics and a microscopic theory. However, a fruitful hydrodynamic description of driven diffusive systems far from equilibrium, the macroscopic fluctuation theory (hereafter MFT) has been proposed \citep{Bertini2015}. It is based on a variational principle which provides equations for the time evolution of the most probable density profile corresponding to a given fluctuation. The MFT was used to explore aspects of out of equilibrium systems \citep{Derrida2007,Aminov2014,Bertini2010a,Bodineau2008,Bertini2009,Krapivsky2012a}. The case of current fluctuations has been singled out due to its relevance to a broad range of problems generically known as full counting statistics which play an important role both in classical and quantum systems  \citep{Kamenev,Karrasch2013,Dhar2008,Bernard2012,Levitov1993}. Quite often, a classical description is convenient enough to account for the behaviour of quantum systems driven out of equilibrium. Noise and current statistics in disordered quantum mesoscopic conductors or wave speckles \citep{Akkermans2007,Zyuzin1987}, non equilibrium spins in superconductors \citep{Quay2013} and thermal transport \citep{Biehs2010,Saito2007} provide important examples of such quantum systems. A great amount of effort has been devoted to the investigation of large current fluctuations since they provide a measure of the likeliness of the system to return to equilibrium. Whereas close to equilibrium, energy and density are almost uniform and can be described within linear response theory, for driving currents far enough from the steady state current, the system may preferably choose non uniform and time-dependent solutions for these observables, very much like a dynamical phase transition. 

To make these considerations more precise, we consider a large system of size $L$ connected for a long time to reservoirs of particles at different densities. It reaches a non-equilibrium steady state with a non vanishing and fluctuating particle current. These fluctuations are characterised by the probability $P_t \left( Q \right)$ for having a number of particles $Q$ flowing through the medium during a time $t$. In the long time limit $ t \rightarrow \infty$, this probability follows a large deviation principle,
\be
\frac{1}{t}\log P_{t}\left(Q\right) \equiv -{\frac{1}{L}}\, \Phi_t \left(J=Q/t\right) \, 
\label{ldf}
\ee
where the large deviation function $\Phi_t$ plays in that situation a role similar to the equilibrium free energy \citep{rkconductance}
Expression (\ref{ldf}) is not an obvious result \citep{Touchette2009}. Moreover, finding an explicit expression for $\Phi_t$ is a difficult optimisation problem. However, a useful and elegant additivity principle (hereafter AP) has been formulated \citep{Bodineau2004} which, by assuming that the optimal current trajectory is time independent, allows to reduce the calculation of $\Phi_t$ to solving a Euler-Lagrange equation. A breakdown of the AP signals the onset of a dynamical phase transition. 
 It is the purpose of this letter to formulate a necessary and sufficient condition for the validity of the AP for boundary driven systems with and without additional uniform external field $E$. This will extend results obtained in previous works \citep{Bertini2006,Appert-Rolland2008,Imparato2009} and allow to discuss the existence and the nature of such transitions. 
 
 Despite the fact that out of equilibrium physics requires new approaches which are different from the familiar thermodynamics concepts, it is intuitively helpful to relate these two situations. A powerful idea to study systems at thermodynamic equilibrium is provided by Le Chatelier principle which states that the net outcome of a fluctuation is to bring the system back to equilibrium, or, in other words, thermodynamic potentials are either concave or convex functions. It is possible, using Onsager relations to extend Le Chatelier principle to systems out of equilibrium. To that purpose, we recall that a system brought out of equilibrium by the application of forces $X_i$ such as temperature or density gradients, behaves diffusively and creates fluxes ${\cal J}_i$ linearly related to the forces, ${\cal J}_i = \sum_j L_{ij} \, X_j $. Forces and their related fluxes are defined such that products ${\cal J}_i \, X_j$ are additive terms in the corresponding entropy creation. The symmetric matrix $L_{ij}$ cannot be determined from thermodynamics but only from a microscopic model. A generalisation of Le Chatelier principle is obtained 
from the expression $\mathfrak{s} = \sum_i {\cal J}_i \, X_i$ of the entropy per unit time. Thus, using the definition of the ${\cal J}_i$'s and the symmetry of the $L_{ij}$'s leads to the  {\it positive} quadratic form,
 \be
 \mathfrak{s} = \sum_{ij} L_{ij} \, X_i \, X_j \, ,
 \label{entropyonsager}
 \ee
 which implies that $L_{mm} \geq 0$. Then, varying the force $X_m$ by $\delta X_m$, we obtain from (\ref{entropyonsager}) that ${\cal J}_m \, \delta X_m \geq 0$, namely the flux and the fluctuation generating it are always of the same sign, so that the response of the system tends to act against the perturbation. This is the content of Le Chatelier principle for non equilibrium and its breakdown signals the possible onset of a phase transition. 

We wish now to implement these ideas using the framework of the MFT. To that purpose and for the sake of simplicity, we restrict our study to one-dimensional systems although generalisations to higher dimensions have been proposed \citep{Akkermans2013}. We consider a lattice gas such that 
 $n_i (t)$, $i\in1,...,L$ denote the time-dependent occupancies of the $L \gg 1$ sites of the system coupled to two reservoirs at its endpoints. The MFT relies on the replacement of the dynamics of the system (either deterministic or not) by a stochastic hydrodynamic equation which  describes correctly the fluctuations of the driven system in the long time and large size limits. The relevant physical quantities are the density $\rho \left( x, \tau \right)$ and the current density $j\left(x,\tau\right)$ of a fluctuating diffusive system, with the scaling $x = i/L$ and $\tau = t/ L^2$. The boundary conditions for the density are time-independent and fixed by $\rho_{L\left(R\right)}$ at the left (right) boundaries $x=0,1$. The general evolution of the system in the presence of an external and uniform field $E$, is described by a stochastic Langevin equation, 
\be j\left(x,\tau\right)=-D\left(\rho\right)\partial_{x}\rho + E \sigma (\rho) + \sqrt{\sigma (\rho)} \, \eta\left(x,\tau\right) \, , \label{le} \ee 
together with the continuity equation $\partial_{\tau}\rho=-\partial_{x}j$. The term $\eta\left(x,\tau\right)$ is a multiplicative Gaussian white noise with zero mean
and variance $\frac{1}{L} \delta\left(x-x'\right)\delta\left(\tau-\tau'\right)$. The $1/L$ factor ensures that the noise term is vanishingly small in the limit of a large system thus allowing the use of saddle point methods. The phenomenological diffusion, $D\left(\rho\right)$, and conductivity (transport) $\sigma\left(\rho\right)$ coefficients may be obtained from the details of the microscopic process. On average, the current is determined by a diffusive Fick term and a term proportional to the applied field $E$ (linear response for a weak enough field). At equilibrium, $j=0$ so that these two contributions balance each other thus providing a relation between $D\left(\rho\right)$ and $\sigma\left(\rho\right)$. Out of equilibrium, the strength of the noise term driven by the dissipative conductivity $\sigma\left(\rho\right)$ is related to the diffusion Fick term by means of a Einstein relation which, as expected, coincides with the relation obtained at equilibrium (fluctuation-dissipation relation) \citep{Gallavotti1995}. This generalises the usual Langevin equation where the strength of the stochastic noise is driven by temperature instead of conductivity. The number of particles $Q$ in (\ref{ldf}), is the integral of the current density,
\be
Q=L^{2}\intop_{0}^{1}dx\,\intop_{0}^{t/L^{2}}d\tau\, j\left(x,\tau\right) \, .
\label{intcurrent} \ee
and the two phenomenological coefficients $D\left(\rho\right)$ and $\sigma\left(\rho\right)$ can be expressed using the first two cumulants of the probability $P_t (Q)$ in (\ref{ldf}). To establish these expressions and from now on we consider the case $E=0$ in (\ref{le}) unless stated otherwise. In the limit $\rho_{R}-\rho_{L}=\Delta\rho\ll1$ of a slightly out of equilibrium system, the steady state average current $\left\langle Q\right\rangle /t$ is obtained from (\ref{intcurrent}) and given by $\left\langle Q\right\rangle /t=-\frac{1}{L}D\left(\rho\right)\Delta\rho$. For $\Delta\rho\rightarrow0$, the  variance of the integrated current is 
$\left\langle Q^{2}\right\rangle _{C}/t=\frac{1}{L}\sigma\left(\rho\right)$.


The probability $P_{t}\left(Q,\rho_{L},\rho_{R}\right)$ is obtained in this framework with the help of a stochastic path integral representation (a.k.a Martin-Siggia-Rose \citep{Martin1973,Kamenev}) \citep{rkJordan}
\begin{equation}
P_{t}\left(\left\{ j,\rho\right\} \right)\sim\exp\left[-L\,\int_{0}^{t/L^{2}}d\tau\,\int_{0}^{1}dx\,\mathcal{L}\right],\label{eq:general prob. definition}
\end{equation}
corresponding to a given current and density trajectories
$\left\{ j\left(x,\tau\right),\rho\left(x,\tau\right)\right\} $. The Lagrangian density $\mathcal{L}\left(\rho,\partial_{x}\rho\right)$ is,
\be
\mathcal{L}=\frac{\left(j+D\left(\rho\right)\partial_{x}\rho\right)^{2}}{2\sigma\left(\rho\right)} \, ,
\label{lagrangian} \ee
and the large deviation function in (\ref{ldf}) rewrites \citep{Bertini2006},
\be
\Phi_{t}\left(J = {\frac{Q}{t}} \right)=\frac{L^{2}}{t}\underset{j,\rho}{\inf}\int_{0}^{t/L^{2}}d\tau\,\int_{0}^{1}dx\,\mathcal{L} \, 
\label{ldf1} \ee
where the minimum is over all density profiles $\rho\left(x,\tau\right)$ and current densities $j\left(x,\tau\right)$ defined in the time interval $0 < \tau < t / L^2$ and which satisfy the continuity equation and the constraint (\ref{intcurrent}), $J = Q / t$. The hard minimisation problem of finding the optimal current trajectory $j\left(x,\tau\right)$ becomes much simpler by assuming the optimal current to be constant, $j\left(x,\tau\right)=J$ (up to a transient where the system adjusts to this optimal solution). This approximation introduced by Bodineau and Derrida \citep{Bodineau2004}, is known as the additivity principle (AP). A spatially constant current implies, through the continuity equation, a stationary density $\rho (x)$, so that the corresponding Lagrangian density obtained from (\ref{lagrangian}) and now denoted $\mathcal{L}_J$ becomes time-independent. Therefore, the AP amounts to replacing $\Phi_{t}\left(J\right)$ in (\ref{ldf1}) by the new large deviation function 
\be
U\left(J\right)=\underset{\rho\left(x\right)}{\inf}\int_{0}^{1}dx\,\mathcal{L}_J \left(\rho(x) , \partial_x \rho(x)\right) \, . \label{eq:LDF within AP}
\ee
It is noteworthy that while a Lagrangian is usually integrated w.r.t time, here time is replaced by the spatial coordinate. Both the approximate large deviation function $U\left(J\right)$ and the “trajectory” of $\rho (x)$, which corresponds to the stationary density profile under the AP assumption, are then obtained from the associated Euler-Lagrange equation $\frac{d}{dx}\frac{\delta\mathcal{L}_J}{\delta\partial_{x}\rho}=\frac{\delta\mathcal{L}_J}{\delta\rho}$. This Lagrangian approach can also be addressed in terms of the corresponding  Hamiltonian formalism, where the Hamiltonian can be shown to be given by \citep{rkGalilean}
\begin{equation}
H\left({\cal P},q\right)=\frac{1}{2m\left(q\right)}\left[\mathcal{P} -eA\left(q\right)\right]^{2}-e^{2}V\left(q\right),\label{eq:Hamiltonian trans}
\end{equation}
provided we define $q=\rho$ and the conjugate momentum $\mathcal{P}=\left[ D \left(q\right)/ \sigma\left(q\right) \right] \, \left(J+D\left(q\right)\partial_{x}\rho\right)$ . The Hamiltonian (\ref{eq:Hamiltonian trans})
describes a single particle of $q$-dependent mass $m\left(q\right)=D^2\left(q\right)/ \sigma\left(q\right)$ and of charge $e=J$ placed in scalar $V\left(q\right)=1/ 2\sigma\left(q\right)$ and "vector" $A\left(q\right)=D\left(q\right) / \sigma\left(q\right)$ potentials. As stressed just before,  space replaces time, namely, time conservation in Hamiltonian systems translates here into a conservation in space of the associated energy $H\left(\mathcal{P},q\right)$, so that the energy is spatially equally distributed. This provides an interesting analogy with thermodynamics where, at equilibrium, the total energy is uniformly distributed in space. Therefore, the AP appears to provide, for out of equilibrium systems, the analog  of a thermodynamic description.

A careful study of the conditions under which the AP is valid thus appears to be  essential, since a breakdown of the AP may signal the onset of a (dynamical) phase transition. This question has been investigated in \citep{Bertini2006} for closed systems with periodic boundary conditions, $\rho (0, \tau ) = \rho (1, \tau)$, and a sufficient condition for the validity of the AP has been given. However, in that case, periodic boundary conditions and the additional conserved quantity $\int_0^1 dx \, \rho(x, \tau )$ simplify the problem. Here, we wish to provide a necessary and sufficient condition for the validity of the AP in  boundary driven systems. This question has also been addressed using a direct stability analysis of the large deviation function against time dependent perturbations \citep{Bodineau2007} but without conclusive results. Here,  to implement this program and address the problem of the stability of the AP solution, we find it more convenient to work with the Legendre transform $\mu\left(\lambda\right)$ of the large deviation function $\Phi_t \left(J\right)$, 
\be 
\mu\left(\lambda\right)=-\frac{1}{L}\underset{J}{\inf}\left\{ \Phi_t \left(J\right)-\lambda J\right\} = \frac{1}{t}\ln\left\langle e^{\lambda Q/L}\right\rangle \, ,
\label{mu} \ee
since this removes the constraint on the integrated current $Q$ \citep{rkCGF}. The notation $\left\langle \,\cdot\,\right\rangle $ accounts for averaging
with respect to $P_{t}\left(Q\right)$ given in (\ref{ldf}). Being cautious about  the corresponding change of boundary conditions, it is possible to relate $\mu ( \lambda )$ to the MFT description by means of the relation,
\be
\left\langle e^{\lambda Q/L}\right\rangle =\int\mathcal{D}q\mathcal{D}p\,\exp\left[-L\int dx\, d\tau\, S \left(x,\tau\right)\right],
\label{action} \ee
where $q$ again stands for the density and $p$ is a Lagrange multiplier associated to the continuity equation \citep{supp}. The
action $S\left(x,\tau\right)$ is given by 
\be
S\left(x,\tau\right)= D\partial_{x}q \, \partial_{x}p-\frac{\sigma}{2}\left(\partial_{x}p\right)^{2}+\left(p-\lambda x\right)\partial_{\tau}q \, .
\label{action2} \ee 
The corresponding equations of motion can be readily obtained from $ \delta S / \delta q = \delta S / \delta p =0$ \citep{Jordan2004}, 
\begin{eqnarray}
 \partial_\tau q = \partial_{x}\left(D\partial_{x}q \right)-\partial_{x}\left(\sigma\partial_{x}p\right)  & &   \nonumber \\
\partial_\tau p = -D \, \partial_{xx}p - {\frac{\sigma'}{2}} \left(\partial_{x}p\right)^{2} \,\,\,\, & &  
\label{eq:AP equation of motion  with field-1}
\end{eqnarray}
where the notations $(D' , \sigma' )$ stand for derivatives w.r.t the density $q$. Now we consider the AP which assumes time-independent density and momentum, so that taking $\partial_\tau q = \partial_\tau p =0$, the AP equations of motion become two ordinary differential equations for the corresponding $\left( q_{0} , p_{0} \right)$ with the boundary time independent conditions

\be
\begin{cases}
q\left(0,\tau \right)=\rho_{L} & q\left(1,\tau \right)=\rho_{R}\\
p\left(0,\tau \right)=0 & p\left(1,\tau \right)=-\lambda \, \, .
\end{cases}
\label{bc} \ee


The most probable density profile under the AP, is obtained by solving these Hamilton-Jacobi  equations with boundary conditions (\ref{bc}). 

In order to discuss the stability of the AP solution, we consider the effect of a time-dependent fluctuation $\delta q (x,\tau)$ and $\delta p (x,\tau)$ of the density and its conjugate momentum on the extremum solution $\left( q_{0} , p_{0} \right)$ and we calculate the variation $\delta S_{AP}^2$ up to second order in $\left( \delta q, \delta p \right)$ of the action (\ref{action2}). The resulting  diagonal quadratic form \citep{rkdsigma}
,
 \begin{equation}
\delta S_{AP}^2 \left( x, \tau \right) =-\frac{D'\sigma'-\sigma''D}{4 D}\left(\partial_{x}p_0 \right)^{2}\delta q^{2}-{\frac{\sigma}{2}} \left(\partial_{x}\delta p\right)^{2} 
 \label{eq:quadratic form d2S-2}
\end{equation}
 which could not be easily anticipated \citep{supp,rkquadratic} 
 constitutes one of the main results of this letter.
   It allows to discuss the validity of the AP approximation and the onset of dynamical phase transitions. The stability of the AP solution requires $\int dx \, d\tau \, \delta S_{AP}^2 < 0$, a condition which corresponds to Le Chatelier condition  (\ref{entropyonsager}), and since $\sigma$ and $D$ are non negative (for any $q$), then, having 
\be 
D'\sigma'\geq\sigma''D \, ,
\label{suff} \ee
implies $\int dx \, d\tau \, \delta S_{AP}^2 \leq 0$ for any fluctuation $\delta q,\delta p$. Therefore (\ref{suff}) is a sufficient condition for validity of the AP solution. A similar condition has been obtained by Bertini \textit{et al.} \citep{Bertini2006}. But here we do not require having (\ref{suff}) for any $q$, but only for the density profile $ q_{0} = \rho_{AP} $ of the AP solution. However, since the variations $\delta q$ and $\delta p$ are not independent but related by (the conveniently linearised) equations (\ref{eq:AP equation of motion  with field-1}), it is clear that (\ref{suff}) is not a necessary condition for stability.


To find a necessary condition for the instability of the AP solution, we consider the Fourier spectrum of the time-dependent fluctuations $\delta q (x,\tau)$ and $\delta p (x,\tau)$. Since time is defined on the interval $[0, T]$ where $T=t/L^{2}$, these fluctuations admit the Fourier series expansion, $\delta q=\sum_{\omega}e^{i\omega\tau}f_{\omega}\left(x\right) $ and $\delta p=\sum_{\omega}e^{i\omega\tau}g_{\omega}\left(x\right) $ with discrete frequencies $\omega_m =\frac{2\pi}{T}m$, $( m \in \mathbb{Z})$. The linearisation of equations (\ref{eq:AP equation of motion  with field-1}) \citep{supp} previously used  to obtain the quadratic form (\ref{eq:quadratic form d2S-2}) together with the condition of real valued fluctuations, lead for the Fourier amplitudes to the set 
of coupled differential linear equations \citep{rkdsigma}
\begin{eqnarray}
i\omega f_{\omega} &=& \partial_{x}\left(D'\partial_{x}q_0 \, f_{\omega}+D \, \partial_{x}f_{\omega}-\sigma' (\partial_{x}p_0 ) \, f_{\omega}-\sigma\partial_{x}g_{\omega}\right)  \nonumber \\
i\omega g_{\omega} &=& \left( -D'\partial_{xx}p_0  - {\frac{\sigma''}{2}} \left(\partial_{x}p_0\right)^{2} \right) f_{\omega}-D\partial_{xx}g_{\omega}  
\nonumber \\ 
&-& \sigma'\partial_{x}p_0 \, \partial_{x}g_{\omega} 
\label{neccond2} 
\end{eqnarray}
which, together with the equalities $\int d\tau\,\delta q^{2}=\sum_{\omega>0}\left|f_{\omega}\right|^{2}$ and $\int d\tau\,\left(\partial_{x}\delta p\right)^{2}=\sum_{\omega>0}\left|\partial_{x}g_{\omega}\right|^{2}$, allow to rewrite the corresponding total fluctuation of the action (\ref{eq:quadratic form d2S-2}) as $\int dx \, d\tau\, \delta S_{AP}^{2}\left(x,\tau\right) =-\sum_{\omega>0}\delta s^2 _{\omega} $, where \citep{rkdsigma}
\be
\delta s^2 _{\omega} \equiv \int dx\,\frac{D'\sigma'-D\sigma''}{4 \, D} 
\left\{ \left(\partial_{x}p_0 \right)^{2} \right\} \left|f_{\omega}\left(x\right)\right|^{2}+{\frac{\sigma}{2}} \left|\partial_{x}g_{\omega}\left(x\right)\right|^{2}  
\label{omega} \ee
Thus, a necessary condition for the AP solution to be unstable against a small time dependent perturbation is the existence of at least one unstable mode $\omega_0$ such that $\delta s^2 _{\omega_0} < 0$. This condition together with the linear set of equations (\ref{neccond2}) give a constructive and easy to implement prescription to find the frequency $\omega_0$ and spatial amplitude $f_{\omega_0}\left(x\right)$ of a time-dependent density mode  $\rho \left(x,\tau\right)=\rho_{AP}\left(x\right)+\left[e^{i\omega_0 \tau}f_{\omega_0}\left(x\right)+e^{-i\omega_0\tau}f_{\omega_0}^{*}\left(x\right)\right]$ which minimises the action (\ref{action2}). Similar considerations applied to systems with spatial periodic boundary conditions \citep{Appert-Rolland2008,Bodineau2005,Espigares2013,Saito2011}, lead to a closed expression of such an unstable mode $\omega_0$. Although a closed  expression can hardly be obtained for open systems considered here, the general conclusions seem to hold in that case as well, namely, for finite size $L$ and long time limit $t \rightarrow \infty$, the first unstable mode is expected to be the fundamental so that the system is driven through a continuous, second order like transition \citep{Appert-Rolland2008}.

The previous considerations extend to the case of an open and boundary driven system in the presence of an additional uniform external field $E$ as described by the stochastic equation (\ref{le}). The corresponding Lagrangian rewrites $\mathcal{L}_{E}=\left(J+D\left(\rho\right)\partial_{x}\rho-E\sigma\left(\rho\right)\right)^{2} / 2\sigma\left(\rho\right)$ instead of (\ref{lagrangian}). The time-independent AP Hamilton-Jacobi equations become 
\begin{eqnarray}
\partial_{x}\left(D\partial_{x}q-E\sigma\right)-\partial_{x}\left(\sigma\partial_{x}p\right) &=& 0 \nonumber \\ 
-D\partial_{xx}p-E\sigma'\partial_{x}p-\frac{\sigma'}{2}\left(\partial_{x}p\right)^{2} &=& 0 
\label{withE} \end{eqnarray} 
with the same boundary conditions (\ref{bc}). These equations are obtained from the modified action $S_{E}\left(x,\tau\right)=\left(D\partial_{x}q-E\sigma\right)\partial_{x}p-\frac{\sigma}{2}\left(\partial_{x}p\right)^{2}+\left(p-\lambda x\right)\partial_{\tau}q 
$ instead of (\ref{action2}). To study the stability of the AP solution, we evaluate, as previously, the variation $\delta S_{E}^2$ up to second order of the AP action under the effect of a fluctuation $\delta q$ of the density and  $\delta p$ of its conjugate momentum. $\delta S_{E}^2$ is again given by the  diagonal quadratic form (\ref{eq:quadratic form d2S-2}) except for the replacement of $\left(\partial_{x}p_0\right)^{2}$ by $\left(\partial_{x}p_0 \right)^{2}+2E\partial_{x}p_0$. Therefore, unlike the case $E=0$, we cannot a priori  conclude that (\ref{suff}) is a sufficient condition for the stability of the AP solution. However, it happens that we indeed always have $\left(\partial_{x}p_0 \right)^{2}+2E\partial_{x}p_0 >0$. This is a consequence of the AP equations (\ref{withE}). 
Defining $u=\partial_{x}p_0 +E$ allows to rewrite the second equation of (\ref{withE}) under the form \citep{rkdsigma}
\begin{equation}
\frac{\partial_{x}u}{u^{2}-E^{2}}=-{\frac{\sigma'}{2D}}.\label{eq: u equation to obtain AP with field}
\end{equation}
Next, we define $h\left(x\right) \equiv \int dx\, {\frac{\sigma'\left(q_0\right)}{2 D\left(q_0\right)}}  $ for a known AP density profile $q_0\left(x\right)$. An  integral of (\ref{eq: u equation to obtain AP with field}) is implicitly obtained in terms of $h(x)$ under the form $u=E\coth\left(E \, h\left(x\right)\right)$. Therefore, $\left(\partial_{x}p_0\right)^{2}+2E\partial_{x}p_0 =E^{2}/\sinh^{2}\left(E \, h\left(x\right)\right)>0$
for any $E$ so that (\ref{suff}), $D'\sigma' \geq D\sigma''$, remains a sufficient condition for stability of the additivity principle solution even in the presence of a field $E$. 

The condition (\ref{suff}) leads to new results. We first consider the simple symmetric exclusion process (SSEP) which constitutes a paradigm for non equilibrium behaviour \citep{Mallick2015,Derrida2004}. In that case, the large deviation function is exactly calculated using either the Bethe ansatz or matrix representation \citep{Derrida2011}. Moreover, the SSEP can also be described using the MFT by means of a dynamics defined by a constant (i.e. $\rho$-independent) diffusion coefficient $D = 1$  and the conductivity $\sigma (\rho) = 2 \rho \, ( 1 - \rho)$. In that case (\ref{suff}) is readily satisfied thus proving in another way the already known stability of the SSEP for a boundary driven process. More involved and an open problem as yet, is the stability of the SSEP under an external uniform field $E$, known as the weakly asymmetric exclusion
process (WASEP), which possess the SSEP dynamics given above with a driving field
\citep{Bodineau2006}. Since the SSEP is stable for a boundary driven process, so is the WASEP \citep{Gorissen2012}. It is nevertheless worth noting that in the case of periodic boundary conditions,
(\ref{suff}) is no longer applicable due to the additional constraint of particle conservation. And indeed for periodic systems, the WASEP was found
to be unstable and certain values of the current lead to travelling wave solutions \citep{Appert-Rolland2008}.

In summary, we have presented a new quantitative approach to study the stability of boundary driven systems out of equilibrium. This approach based on the stochastic MFT, provides a necessary and sufficient condition expressed by (\ref{suff}) for stability of the AP solution and the onset of a dynamical phase transition. It constitutes a generalisation of Le Chatelier stability principle. Moreover, in that framework, we have been able to prove the stability of the (boundary driven) WASEP model. 

Another well studied model is the KMP model \citep{Kipnis1982}, which corresponds to $D=1$ and $\sigma = 2\rho^2$. Clearly, the KMP model does not satisfy (\ref{suff}) thus being non conclusive about its stability. However, solving numerically (\ref{neccond2}) for a large range of currents \citep{rkconj}
  seems to preserve the AP stability. Those results can be justified on the basis of scaling arguments \citep{rkslava}. 
  This suggests that the KMP model should also be stable for boundary driven systems, in agreement with \citep{Hurtado2009}. An important asset of the MFT approach and of our stability analysis resides in their potential relevance to a class of problems larger than used so far, such as quantum mesoscopic transport of particles \citep{Zyuzin1987}, classical waves in a random potential (speckle correlations) including higher order quantum corrections \citep{Akkermans2007}, thermal conductance in quantum chains \cite{Karrasch2013}, cold atoms \citep{Esslinger} and polarised spins injected into superconductors \citep{Quay2013}. This approach may also be relevant and shed light on the problem of quantum thermal transport and its relation with conformal description \citep{Bernard2012}.

Acknowledgements: This work was supported by the Israel Science Foundation Grant No.924/09. E.A. wishes to thank the College de France (Paris) for the support of a Chaire d'Etat and O.S. support by Ecole Polytechnique.  E.A and O.S. acknowledge fruitful discussions with K. Mallick, T. Bodineau and D. Bernard.


\begin{thebibliography}{99}


\bibitem{Redner} P. L. Krapivsky , S. Redner and E. Ben-Naim  {\it A Kinetic
View of Statistical Physics}, (Cambridge
University Press, 2010). 

\bibitem{Bertini2015} L. Bertini, A. De Sole, D. Gabrielli, G. Jona-Lasinio, and C. Landim, Rev. Mod. Phys. 87, 593 (2015).

\bibitem{Derrida2007} B. Derrida, Journal of Statistical Mechanics: Theory and Experiment 2007, P07023 (2007).

\bibitem{Aminov2014} A. Aminov, G. Bunin, and Y. Kafri, Journal of Statistical Mechanics: Theory and Experiment 2014, P08017 (2014).

\bibitem{Bertini2010a} L. Bertini, A. De Sole, D. Gabrielli, G. Jona-Lasinio, and C. Landim, Journal of Statistical Mechanics: Theory and Experiment 2010, L11001 (2010).

\bibitem{Bodineau2008} T. Bodineau, B. Derrida, V. Lecomte, and F. van Wijland, J. Stat. Phys. 133, 1013 (2008). 

\bibitem{Bertini2009} L. Bertini, A. De Sole, D. Gabrielli, G. Jona-Lasinio, and C. Landim, J. Stat. Phys. 135, 857 (2009).

\bibitem{Krapivsky2012a} P. L. Krapivsky, B. Meerson, and P. V. Sasorov, Journal of Statistical Mechanics: Theory and Experiment 2012, P12014 (2012).

\bibitem{Kamenev} A. Kamenev, {\it Field theory of non-equilibrium systems} (Cambridge University Press, 2011).

\bibitem{Karrasch2013} C. Karrasch, R. Ilan, and J. E. Moore, Phys. Rev. B 88, 1 (2013).

\bibitem{Dhar2008} A. Dhar, Advances in Physics 57, 457 (2008).

\bibitem{Bernard2012} D. Bernard and B. Doyon, J. Phys. A 45, 362001 (2012).


\bibitem{Levitov1993} L. S. Levitov and G. B. Lesovik, Jetp Lett. 58, 230 (1993). L. S. Levitov, H. Lee, and G. B. Lesovik, J. Math. Phys. 37, 4845 (1996).

\bibitem{Akkermans2007}  E. Akkermans and G. Montambaux,
{\it Mesoscopic Physics of Electrons and Photons}, Chap.12, 
(Cambridge University Press, 2007).



\bibitem{Zyuzin1987} A.Yu. Zyuzin and B.Z. Spivak, Sov. Phys. JETP 66 (3), 560 (1987) and R. Pnini and B. Shapiro, Phys. Rev. B {\bf 39} 6986 (1989).
\bibitem{Quay2013} C. H. L. Quay, D. Chevallier, C. Bena, and M. Aprili, Nature Physics 9, 84 (2013).

\bibitem{Biehs2010} S. A. Biehs, E. Rousseau, and J.-J. Greffet, Phys. Rev. Lett. 105, 234301 (2010).


\bibitem{Saito2007} K. Saito and A. Dhar, Phys. Rev. Lett. 99, 180601 (2007).



\bibitem{rkconductance} The $1/L$ dependence holds in one-dimensional systems. More generally, the large deviation function is a conductance and therefore scales analogously, namely as $1/ L^{d-2}$ in $d$-dimensions.




\bibitem{Touchette2009} H. Touchette, Physics Reports 478, 1-69 (2009) 

\bibitem{Bodineau2004} T. Bodineau and B. Derrida, Phys. Rev. Lett. 92, 180601 (2004).

\bibitem{Bertini2006} L. Bertini, A. De Sole, D. Gabrielli, G. Jona-Lasinio, and C. Landim, J. Stat. Phys. 123, 237 (2006). 

\bibitem{Appert-Rolland2008} C. Appert-Rolland, B. Derrida, V. Lecomte, and F. van Wijland, Phys. Rev. E 78, 021122 (2008).



\bibitem{Imparato2009} A. Imparato, V. Lecomte, and F. van Wijland, Phys. Rev. E 80, 011131 (2009).


\bibitem{Akkermans2013} E. Akkermans, T. Bodineau, B. Derrida, and O. Shpielberg, Europhys. Lett. 103, 20001 (2013).

\bibitem{Gallavotti1995} G. Gallavotti and E. Cohen, Phys. Rev. Lett. 74, 2694 (1995). 



\bibitem{Martin1973} P. C. Martin, E. Siggia, and H. Rose, Physica A: Statistical Mechanics and its Applications 8, 423 (1973).
\bibitem{rkJordan} A similar stochastic path integral has been proposed in \citep{Jordan2004} for a classical description of quantum mesoscopic systems. Yet, the analogy is still incomplete, e.g., AP and dynamical phase transition have not been discussed in that framework.

\bibitem{rkGalilean} This canonical form is the one required by Galilean invariance.


\bibitem{Bodineau2007} T. Bodineau and B. Derrida, C. R. Physique 8, 540 (2007).


\bibitem{rkCGF} From that expression, it is clear that $\mu (\lambda)$ is the cumulant generating function of the current. 

\bibitem{supp} Supplemental material.


\bibitem{Jordan2004} S. Pilgram, A. N. Jordan, E. V. Sukhorukov, and M. B{\"u}ttiker, Phys. Rev. Lett. 90, 206801 (2003) and A. N. Jordan, E. V. Sukhorukov, and S. Pilgram, Journal of Mathematical Physics 45, 4386 (2004). 

\bibitem{rkdsigma} In that expression, the phenomenological coefficients $D$ and $\sigma$ and their derivatives are evaluated at $\rho = q_0$. 

\bibitem{rkquadratic} Nevertheless, it is worth noting that the resulting diagonal quadratic form could have been guessed on the basis of the equivalent Hamiltonian description (see \cite{supp} and O. Shpielberg (unpublished)).


\bibitem{Bodineau2005} T. Bodineau and B. Derrida, Phys. Rev. E 72, 066110 (2005).

\bibitem{Espigares2013} C. Espigares, P. Garrido, and P. Hurtado, Phys. Rev. E 87, 032115 (2013).

\bibitem{Saito2011} K. Saito and A. Dhar
Phys. Rev. Lett. 107, 250601 (2011).


\bibitem{Mallick2015} K. Mallick, Physica A: Statistical Mechanics and its Applications 418, 17 (2015).

\bibitem{Derrida2004} B. Derrida, B. Dou{\c c}ot, and P.-E. Roche, J. Stat. Phys. 115, 717 (2004).

\bibitem{Derrida2011} B. Derrida, Journal of Statistical Mechanics: Theory and Experiment 2011, P01030 (2011).

\bibitem{Bodineau2006} T. Bodineau and B. Derrida, J. Stat. Phys. 123, 277 (2006).

\bibitem{Gorissen2012} M. Gorissen and C. Vanderzande, Phys. Rev. E 86, 051114 (2012).

\bibitem{Kipnis1982} C. Kipnis, C. Marchioro, and E. Presutti, J. Stat. Phys. 27, 65 (1982).

\bibitem{rkconj} Here the current is represented by its conjugate $\lambda$. Even though $\lambda$ does not appear explicitly in (\ref{neccond2}), it enters the AP solution $(q_0,p_0)$ through the boundary conditions (\ref{bc}).

\bibitem{rkslava}  O. Shpielberg and Y. Don  (unpublished results).

\bibitem{Hurtado2009} P. Hurtado and P. Garrido, Phys. Rev. Lett. 102, 250601 (2009).


\bibitem{Esslinger}  J. - P. Brantut, J. Meineke, D. Stadler, S. Krinner, and T. Esslinger. Science 337, 1069-1071 (2012). 




\end{thebibliography}

%
%
%

%
%
%

\renewcommand{\thefootnote}{\alph{footnote}}

\end{document}


\title{Supplemental material for "Le Chatelier principle for out of equilibrium and boundary driven systems : application to dynamical phase transitions"}


\author{O. Shpielberg $^{1}$ and E. Akkermans$^1$}
\affiliation{$^1$Department of Physics, Technion Israel Institute of Technology, Haifa 32000, Israel}

\begin{abstract}

In this Supplemental Material, we derive in detail some of the results mentioned in "Le Chatelier principle for out of equilibrium and boundary driven systems: application to dynamical phase transitions". First, we show how to obtain the cumulant generating function formalism. Second, we show how to obtain the diagonal form (15) in the text. We note that this approach can be taken to probe the stability of any Hamiltonian system, in or out of equilibrium. Note, a referenced equation below directs to the equation in the supplemental material unless written otherwise. The referenced citations always refer to the main text.
\end{abstract}

\maketitle

\section{Cumulant generating function}

This section is devoted to explaining (11), namely
\begin{equation}
\left\langle e^{\lambda Q/L}\right\rangle =\int\mathcal{D}q\mathcal{D}p\,\exp \left[ -L\int dx\, d\tau\, S\left(x,\tau\right) \right] \label{exp_lam}
\end{equation}
of the main text. By definition of the large deviation function given in (5) of the main text, we have 
\[
\left\langle e^{\lambda Q/L}\right\rangle =\int\mathcal{D}j\mathcal{D}\rho\,\delta\left(\partial_{\tau}\rho+\partial_{x}j\right)\exp \left[ -L\int dx\, d\tau\,\left\{ \frac{\left[j+D\left(\rho\right)\partial_{x}\rho\right]^{2}}{2\sigma\left(\rho\right)}-\lambda j\right\} \right] \quad ,
\]
with the integration being over all possible currents $j$ and density profiles $\rho$
that satisfy the continuity equation. We introduce a Lagrange multiplier $\tilde{\rho}$ to remove the constraint resulting from the continuity equation.
\[
\left\langle e^{\lambda Q/L}\right\rangle =\int\mathcal{D}j\mathcal{D}\rho\mathcal{D}\tilde{\rho}\,\exp \left[ -L\int dx\, d\tau\,\left\{ \frac{\left[j+D\left(\rho\right)\partial_{x}\rho\right]^{2}}{2\sigma\left(\rho\right)}-\lambda j+\tilde{\rho}\left(\partial_{t}\rho+\partial_{x}j\right)\right\} \right].
\]
Integrating by parts the term $\tilde{\rho}\, \partial_x j $ and completing the square enables to  integrate with respect to $\mathcal{D}j(x,\tau)$ a Gaussian integral. This gives (\ref{exp_lam}),
where $p=\tilde{\rho}-\lambda x$ and $q=\rho$ and the action $S\left(x,\tau\right)$ is
\[
S\left(x,\tau\right)=D\partial_{x}q\partial_{x}p-\frac{1}{2}\sigma\left(\partial_{x}p\right)^{2}+\left(p-\lambda x\right)\partial_{\tau}q,
\]
together with the boundary conditions,
\[
\begin{cases}
q\left(0,\tau\right)=\rho_{L} & q\left(1,\tau\right)=\rho_{R}\\
p\left(0,\tau\right)=0 & p\left(1,\tau\right)=-\lambda \, .
\end{cases}
\]
The macroscopic limit $L\gg1$, allows to make a saddle point approximation for $<e^{\lambda Q / L}>$. The cumulant generating function given in (10) of the main text, is  $\mu\left(\lambda\right)=-\frac{L}{t}\,\underset{q,p}{\inf}\int dx\, d\tau\, S\left(x,\tau\right)$.
The minimization yields (much in the sense of a Hamiltonian formalism)
two equations of motion for the density profile $q$ and its conjugate
momentum $p$
\begin{equation}
\begin{array}{c}
\partial_{\tau}q=\partial_{x}\left(D\partial_{x}q -\sigma\partial_{x}p\right) \\
\quad \quad \partial_{\tau}p=-D\partial_{xx}p-\frac{1}{2}\sigma'\left(\partial_{x}p\right)^{2}.
\end{array}
\label{eq:time dep EoM, no field }
\end{equation}
We note that the first equation in (\ref{eq:time dep EoM, no field }) is simply a restatement of the continuity equation, $\partial_\tau q = - \partial_x j$, for $j=\sigma \partial_x p - D \partial_x q$. This will become important in the derivation the diagonal form (15) of the main text (next section).

\section{General approach for the stability of Hamiltonian systems}

This section is devoted to establishing the diagonal quadratic form given by (15) in the main text. First, we develop a general approach for stability of Hamiltonian systems. Then, we apply it, to the particular case of interest here, namely (15) in the main text. Let us define $s(q,p)$ by
\begin{equation}
s = \int dx d\tau \, \left(	p \partial_\tau	q- \mathcal{H}\right) \quad, \label{eq: the action as a function of the H density}
\end{equation} 
where 
 $\mathcal{H} (p,q)$ is a Hamiltonian density. The minimization of $s$ is found for $(q,p)$ that satisfy $\frac{\delta s}{\delta q}=0$ and $\frac{\delta s}{\delta p}=0$. From (\ref{eq: the action as a function of the H density}), it is easy to see that these two equations yield Hamilton-Jacobi equations are 
\begin{equation} 
\begin{array}{c}
\partial_{\tau}q=\,\,\frac{\delta\mathcal{H}}{\delta p} \quad \, \, \\
\,\, \partial_{\tau}p=-\frac{\delta\mathcal{H}}{\delta q} \quad . 
\end{array} \label{eq: Hamiltons eq}
\end{equation} 
Any $(q,p)$ satisfying Hamilton-Jacobi equations are only guaranteed to be an extremal solution. To ensure a minimal solution, we must require the convexity of $s$ for any small perturbation $(\delta q, \delta p)$ around the extremal solution $(q,p)$, namely, 
\[
s(q+\delta q,p+\delta p)-s(q-\delta q,p-\delta p) - 2s(q,p) >0 \,.
\] 
Hereafter, we evaluate the derivatives of $\mathcal{H}$ at the extremal solution $(q,p)$. For small fluctuations, we have $\delta s^2 >0$ where 
\begin{equation} 
\delta s^2 = \int dx d\tau \,
  ( \frac{1}{2}\frac{\delta^2 \mathcal{H}}{\delta q^2}\delta q^2 +\frac{\delta^2 \mathcal{H}}{\delta q \delta p} \delta p \delta q +\frac{1}{2}\frac{\delta^2 \mathcal{H}}{\delta p^2}\delta p^2 -\delta p \partial_\tau \delta q)  \, .
 \label{eq: delta2s}
\end{equation} 
As noted in the previous section, the continuity equation (here $\partial_\tau q  = \delta \mathcal{H} / \delta p$) implies that the perturbations are not independent. We now linearise this continuity equation around the extremal solution $(q,p)$, i.e.,
\begin{equation}
\partial_\tau \delta q = \frac{\delta^2 \mathcal{H}}{\delta p^2}\delta p +\frac{\delta^2 \mathcal{H}}{\delta p \delta q}\delta q . \label{eq: linearized cont. eq.}
\end{equation}
Inserting it in (\ref{eq: delta2s}) gives 
\begin{equation} 
\delta s^2 = \frac{1}{2} \int dx d\tau (\frac{\delta^2 \mathcal{H}}{\delta q^2}\delta q^2 - \frac{\delta^2 \mathcal{H}}{\delta p^2}\delta p^2) \, . \label{eq: linearised form}
\end{equation}  
We stress again that $(\delta q, \delta p)$ are coupled. To obtain their explicit expressions one must solve the set of linearised equations 
\[
\begin{array}{c}
\partial_{\tau}\delta q=\frac{\delta^{2}\mathcal{H}}{\delta p\delta q}\delta q+\frac{\delta^{2}\mathcal{H}}{\delta p^{2}}\delta p \,\, \\
\\
\,\,\,\, \partial_{\tau}\delta p=-\frac{\delta^{2}\mathcal{H}}{\delta q^{2}}\delta q-\frac{\delta^{2}\mathcal{H}}{\delta p\delta q}\delta p \, .
\end{array}
\]
It is useful to notice that this overall approach does not require the extremal solution to be time independent (as in the AP). Hence, this approach is useful for stability analysis of any action $s$ of the form (\ref{eq: the action as a function of the H density}). 


%
We now apply the previous considerations to the specific case of the MFT theory developed in the letter. In that case, we wish to choose a Hamiltonian density $\mathcal{H}$ such that the action $s$ in (3) corresponds to the action $\int d\tau dx S(x, \tau)$ given by (12) in the main text. 

A possible choice is $\mathcal{H} = - D \partial_x q \partial_x p + \frac{1}{2} \sigma (\partial_x p)^2.$
To complete the analogy, the extra term $ - \lambda x \partial_\tau q$  can be shown to vanish identically when calculating the second order variation. Moreover, that extra term can be safely taken to zero from the beginning  in the Hamiltonian density, since its contribution to the total action is at most constant. Therefore, its contribution to the cumulant generating function $\mu$ vanishes like $ 1/ t$ in the large time limit. 

The Hamilton-Jacobi equations (\ref{eq: Hamiltons eq}) corresponds to (\ref{eq:time dep EoM, no field }). A second order calculation together with integration by parts, leads to
\[ 
\delta s^2 = \int dx d\tau \delta S^2_{AP},
\]
where $\delta S^2_{AP}$ is given by (15) in the main text. 

The generalization to the Hamiltonian density $\mathcal{H}$ to the case of an external field $E$ leads to the expression
\[
\mathcal{H}_E = - (D\partial_x q )\partial_x p - \frac{1}{2}\sigma (  (\partial_x p)^2+ 2E\partial_x p )
\]
of the Hamiltonian density with the corresponding Hamilton-Jacobi equation,

\begin{eqnarray}
\partial_{\tau}q & = & \partial_{x}\left(D\partial_{x}q-E\sigma\right)-\partial_{x}\left(\sigma\partial_{x}p\right)  \\
\partial_{\tau}p & = & -D\partial_{xx}p-E\sigma'\partial_{x}p-\frac{1}{2}\sigma'\left(\partial_{x}p\right)^{2} \, , \nonumber
\end{eqnarray}
where the derivatives of D and $\sigma$ are taken with respect to $q$. Moreover, $D,\sigma$ and their derivatives are evaluated at $q$, the extremal solution.
Linearising around the extremal solution, gives, 

\begin{eqnarray}
\partial_{\tau}\delta q & = & \partial_{x}\left(D'\partial_{x}q\delta q+D\partial_{x}\delta q-E\sigma'\delta q-\sigma'\partial_{x}p\delta q-\sigma\partial_{x}\delta p\right) \label{eq:linearized EOM with field}\\
\partial_{\tau}\delta p & = & (-D'\partial_{xx}p-E\sigma''\partial_{x}p-\frac{1}{2}\sigma''\left(\partial_{x}p\right)^{2})\delta q\quad\quad\quad\quad\quad\quad\quad\quad\quad \nonumber\\
 & - & \left(E\sigma'+\sigma'\partial_{x}p\right)\left(\partial_{x}\delta p\right)-D{}\partial_{xx}\delta p \quad. \nonumber\quad\quad\quad\quad\quad\quad\quad\quad\quad\quad\quad\quad 
\end{eqnarray}
By inserting the first equation for $\partial_\tau q$ in (\ref{eq:linearized EOM with field}) into (\ref{eq: linearised form}), we obtain the corresponding diagonal form for $E\neq 0$, namely,
\[
\delta s^2 _E = \int dx d\tau	\delta S^2_E(x,\tau) \quad ,	
\]
with
\begin{equation}
\delta S^2_E \left(x,\tau\right)=-\frac{1}{4}\frac{D'\sigma'-D\sigma''}{D}\left\{ \left(\partial_{x}p\right)^{2}+2E\partial_{x}p\right\} \delta q^{2}-\frac{\sigma}{2}\left(\partial_{x}\delta p\right)^{2}.\label{eq:diagonal form}
\end{equation}
This is the diagonal form discussed in the main text. Note that the expressions of $\delta S^2_{AP}$ and $\delta S^2_{E}$ differ only by the replacement of $(\partial_x p)^2$ by $(\partial_x p)^2 + 2E\partial_x p$.
Fourier transforming (\ref{eq:linearized EOM with field}) and making $E=0$ leads to (17) in the main text.